\documentclass[epj,twocolumn]{webofc}
\usepackage[varg]{txfonts}   
\usepackage{amsmath}
\newcommand{\alphas}{\alpha_{\rm s}}
\newcommand{\alphasmZ}{\alphas(\rm m^2_{_{\rm Z}})}
\newcommand{\sqrts}{\sqrt{\rm s}}

\newcommand{\lqcd}{\Lambda_{_{\rm QCD}}}

\newcommand{\epem}{e^+e^-}
\newcommand{\meff}{\rm m{_{\rm eff}}}

\def\mean#1{\ensuremath{\left<#1\right>}}

\woctitle{XLIV International Symposium on Multiparticle Dynamics}
\begin{document}
%
\title{Determination of $\alphas$ at NLO*+NNLL from a global fit of 
the low-$z$ parton-to-hadron fragmentation functions in $\epem$ and DIS collisions}
%
%

\author{Redamy P\'erez-Ramos\inst{1,2,3}\fnsep\thanks{\email{perez@lpthe.jussieu.fr}} \and
        David d'Enterria\inst{4}\fnsep\thanks{\email{dde@cern.ch}}
}

\institute{Sorbonne Universit\'e, UPMC Univ Paris 06, UMR 7589, LPTHE, F-75005, Paris, France
\and       
           CNRS, UMR 7589, LPTHE, 
UPMC Univ. Paris 06, BP 126, 4 place Jussieu, F-75252 Paris Cedex 05, France
\and
           Department of Physics, P.O. Box 35, FI-40014 University of Jyv\"askyl\"a, Jyv\"askyl\"a, Finland 
\and
           CERN, PH Department, CH-1211 Geneva 23, Switzerland
          }

\abstract{%
The QCD coupling $\alphas$ is determined from a combined analysis of experimental $\epem$ and 
$e^\pm$p jet data confronted to theoretical predictions of the energy evolution of the parton-to-hadron 
fragmentation functions (FFs) moments --multiplicity, peak, width, skewness-- at low fractional 
hadron momentum $z$. The impact of approximate next-to-leading order (NLO*) corrections plus 
next-to-next-to-leading log (NNLL) resummations, compared to previous LO+NLL calculations, is discussed. 
A global fit of the full set of existing data, amounting to 360 FF moments at collision energies
$\sqrts\approx$~1--200~GeV, results in $\alphasmZ$~=~0.1189$^{+0.0025}_{-0.0014}$ at the Z mass.
}
\maketitle
\section{Introduction}
\label{intro}

As a consequence of asymptotic freedom in quantum chromodynamics (QCD), 
the strong coupling $\alphas$ decreases logarithmically with increasing
energy scale Q. At leading order, $\alphas\propto\ln^{-1}(\rm Q^2/\lqcd^2)$ 
starting from a value $\lqcd\!\!\approx\,$0.2~GeV where the perturbatively-defined coupling 
diverges and the relevant degrees of freedom are not quarks and gluons (collectively called partons)
but colour-neutral hadrons. Theoretical calculations of the parton energy evolution, through gluon radiation and 
quark-antiquark splitting, usually start at scales well above Q$_0\!\!\approx\,$~1~GeV, i.e. for Q~$\gg \rm Q_0\geq\lqcd$, 
such that perturbation theory can be safely applied as a convergent expansion in powers of $\alphas$,
while softer large-distance ($\rm Q\leq Q_0$) phenomena, including the final hadronization, 
are encoded into experimentally-measured parton-to-hadron fragmentation functions (FFs). 
FFs can be interpreted as the probability for a parton to produce a hadron which 
carries a fraction $z$ of the total longitudinal momentum of the jet.\\

\begin{figure*}[htpb!]
\centerline{
\includegraphics[width=0.505\linewidth]{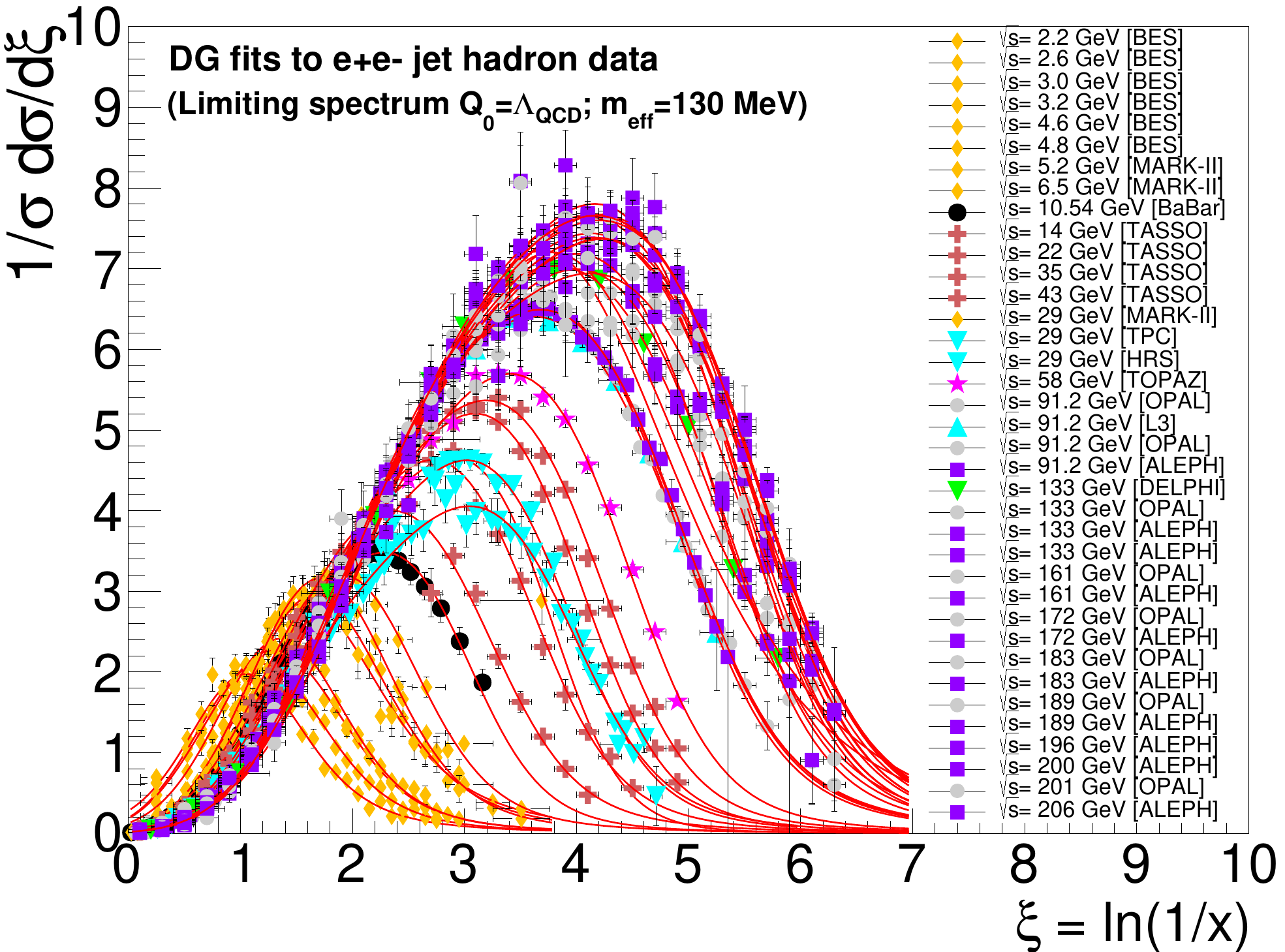}
\includegraphics[width=0.495\linewidth]{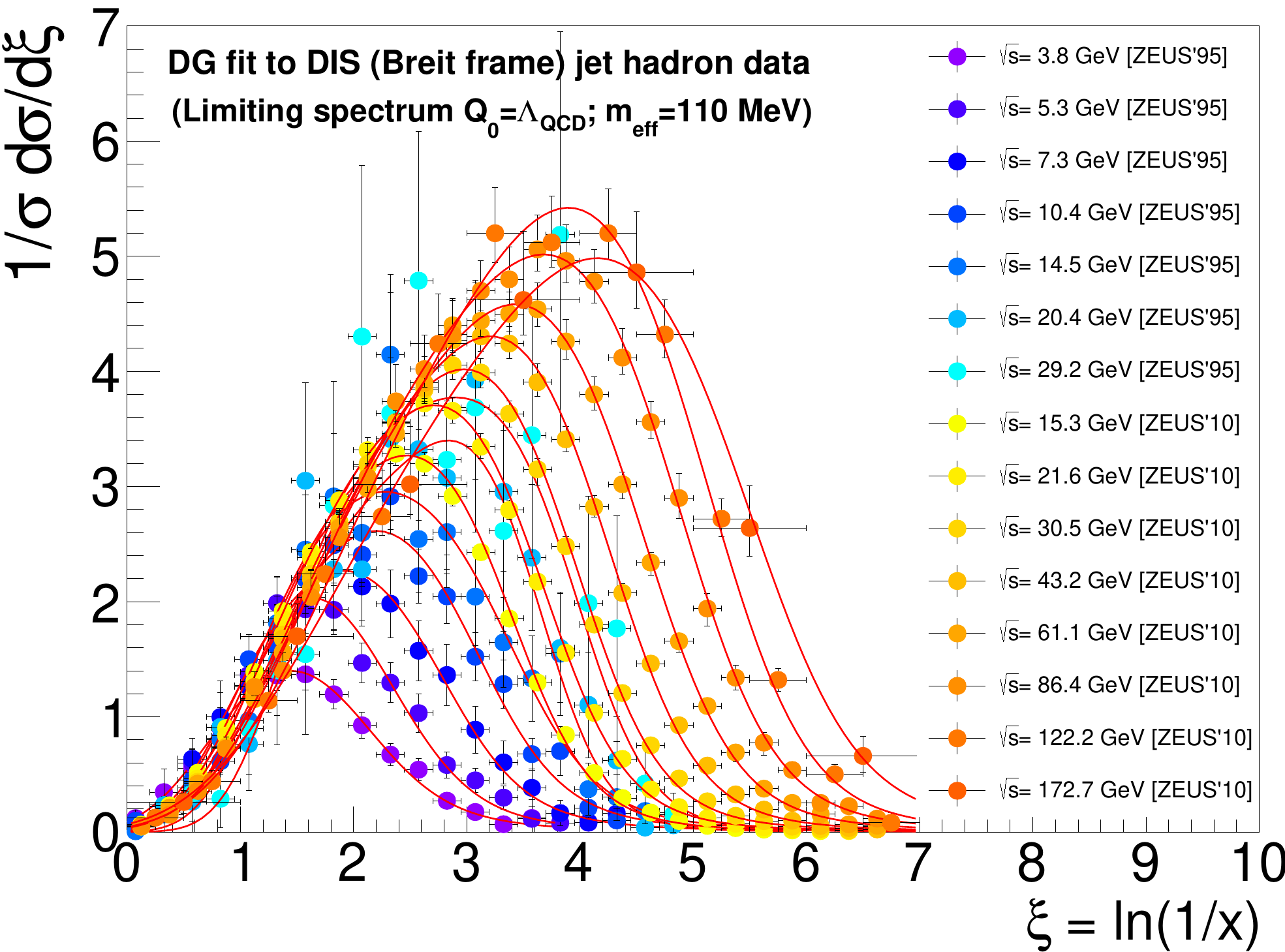}
}
\centerline{
\includegraphics[width=0.505\linewidth]{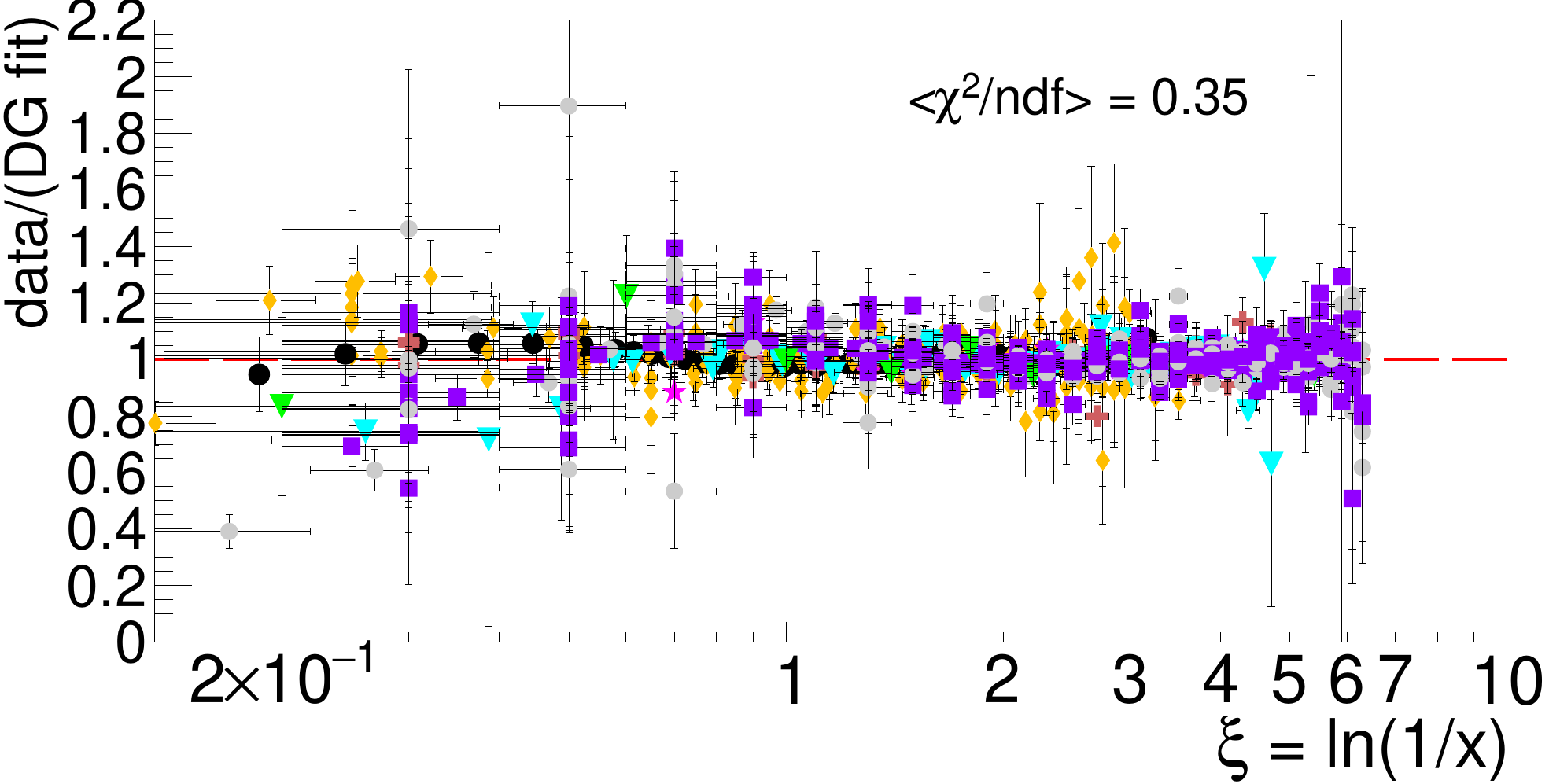}
\includegraphics[width=0.495\linewidth,height=4.28cm]{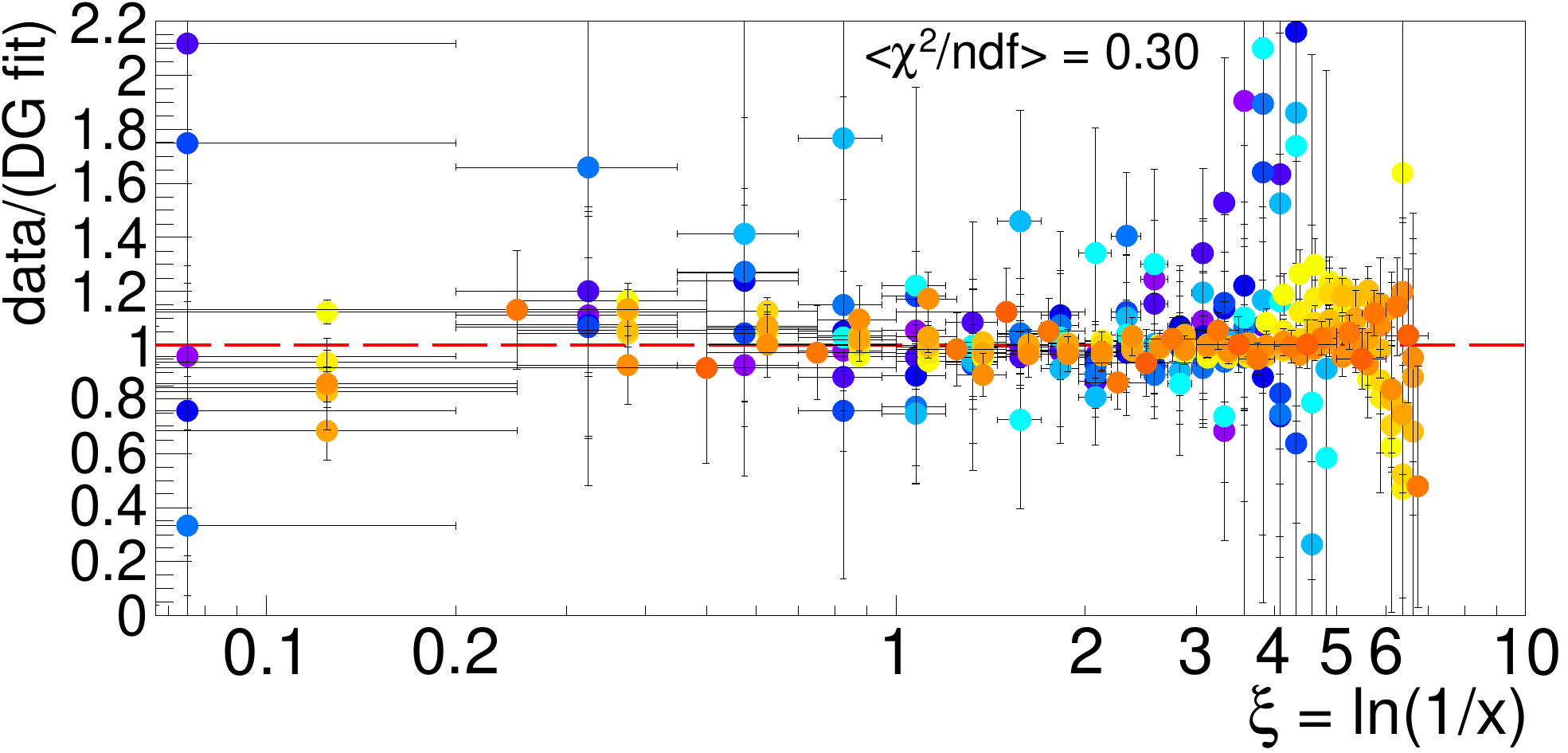}
}
\caption[]{
Top: Charged-hadron distributions in jets as a function of $\xi=\ln(1/z)$ measured 
in $\epem$ at $\sqrts\approx$~2--200~GeV (left) and
$e^\pm,\nu$-p (Breit frame, scaled up by $\times$2 to account for the full
hemisphere) at $\sqrts\approx$~1--180~GeV (right),
individually fitted to the distorted Gaussian, Eq.~(\ref{eq:DG}), with the hadron 
mass corrections ($\meff$~=~120~$\pm$~20~MeV) quoted.
Bottom: Ratio of the measured FFs to each DG fit.
}
\label{fig:DGfits}
\end{figure*}

At large $z\gtrsim$~0.1, one can extract a value of $\alphas$ from the scaling violations of the FFs
at next-to-leading order (NLO) 
accuracy via the comparison of inclusive cross-sections for hadron production measured in high-energy particle
collisions with theoretical predictions~\cite{Kniehl:2000cr}. 
The obtained $\alphasmZ$~=~0.1170$\pm$0.0073 value at the Z mass pole in such approaches is consistent with
the current (NNLO) world average $\alphasmZ$~=~0.1185$\pm$0.0006~\cite{PDG}, derived from a variety of
measurements at $\epem$, deep-inelastic scattering (DIS) e$^\pm$,$\nu$-p, and hadron-hadron colliders. 
The bulk of hadron production from jets is, however, concentrated at low $z\lesssim$~0.1 where
parton evolution is dominated by singularities due to soft and collinear gluon bremsstrahlung~\cite{Dokshitzer:1991wu} 
approaching $\lqcd\approx$~0.2~GeV.
Indeed, due to colour coherence and gluon-radiation interference inside a parton shower,
partons with intermediate energies ($E_h\propto E_{\rm jet}^{0.3}$) 
multiply most effectively in QCD cascades, leading to a final hadron spectrum peaked at low $z$,
with a typical ``hump-backed plateau'' (HBP) shape as a function of $\xi=\ln(1/z)$. 
The HBP shape of the single-inclusive distribution of hadrons in jets can be parametrized, without any loss of
generality, as a distorted Gaussian (DG) which depends on the original energy of the parton, 
${\rm Y} \approx \ln{E/\rm Q_{0}}$, evolved down to a shower cutoff scale $\lambda = \ln(\rm Q_{0}/\lqcd)$:
\begin{equation}
D^{+}(\xi,\rm Y,\lambda) = \frac{{\cal N}}{(\sigma\sqrt{2\pi})}\cdot e^{\left[\frac18k-\frac12s\delta-
\frac14(2+k)\delta^2+\frac16s\delta^3+\frac1{24}k\delta^4\right]}\,, 
\label{eq:DG}
\end{equation}
where $\delta=(\xi-\bar\xi)/\sigma$, with moments:
${\cal N}$ (hadron multiplicity inside the jet), $\bar\xi$ (DG peak position), 
$\sigma$ (DG width), $s$ (DG skewness), and $k$ (DG kurtosis).
Figure~\ref{fig:DGfits} shows the FFs measured in $\epem$ 
(37 datasets~, left) and DIS (15 datasets, right) fitted to the DG expression (\ref{eq:DG}).\\
%

Thanks to kinematical constraints on the parton branching process, such as exact angular 
ordering in the $s$-channel, it is possible to study the evolution of quark and gluon FFs 
also at low $z$ (i.e. at high $\xi$) via renormalized equations that resum all singularities at 
leading-log (LLA) accuracy and beyond (Modified Leading Logarithmic Approximation MLLA~\cite{Dokshitzer:1991ej}, 
and next-to-MLLA~\cite{PerezRamos:2007cr}). 
Attempts to extract $\alphas$ from the low-$z$ (i.e. high-$\xi$) moments of the FFs were
carried out in the past at MLLA accuracy (see e.g.~\cite{Akrawy:1990ha}).
However, these older calculations relied on a number of simplifying assumptions: 
(i) ad hoc cuts in the experimental distributions, (ii) simple fits in a restricted FF range, 
(iii) number of quark-flavours fixed to $N_f$~=~3, (iv) use of only one or two FF
moments (Gaussian approximation), and (iv) LO expression for $\alphas$.
As a result, inconclusive values of $\lqcd\approx$~80--600~MeV were reported (see e.g.~\cite{AlbinoKniehl}), 
although the latest studies of the energy evolution of the first FF moment alone (i.e. of the hadron
multiplicity in jets) have yielded more precise $\alphas$ results consistent with the world average~\cite{Kotikov:2014kda}.
Recently, we have presented a novel extraction of $\alphas$ based on the fit of the energy evolution of the
first four moments of the low-$z$ parton-to-hadron FFs including resummations of next-to-next-to-leading
logarithms (NNLL or NMLLA) complemented with NLO* running-coupling corrections\footnote{The asterisk in the
term 'NLO*' stands for 'approximate NLO' as there are missing corrections in the  splitting functions.}, where more terms
have been consistently added to the perturbative expansion compared to previous works, and where a large set
of experimental FF data has been systematically analysed for the first 
time~\cite{Perez-Ramos:2013eba,d'Enterria:2014yya,d'Enterria:2014bsa}.
Thus, by fitting  the experimental single-inclusive hadron distribution for jets at various energies to the DG parametrization
(\ref{eq:DG}), one can determine $\alphas$ from the corresponding energy-dependence of its fitted moments.
Since the current world-average $\alphas$ uncertainty is of order $\pm$0.5\% --although more conservative estimates
place it at the $\pm$1\% level~\cite{Altarelli:2013bpa}, making of $\alphas$ the least precisely-known of all
fundamental couplings in nature-- having at hand extra independent approaches to determine $\alphas$,
with experimental and theoretical uncertainties different than those of the methods currently
used, is an obvious advantage.\\

In this work, we present first a more detailed study of the relative role of the higher order 
(NLO*+NNLL) corrections included 
in our analytical expressions for the multiplicity, maximum peak position, width, skewness, and  kurtosis of
the FFs, compared to the LO+NLL (or MLLA) expressions obtained in the past. 
In a second stage, we do a combined study of $\epem$ and DIS jet FF data, including a few (older) datasets
not incorporated in our previous analyses~\cite{Perez-Ramos:2013eba,d'Enterria:2014yya,d'Enterria:2014bsa}, and
we carry out a {\it single} global fit of all DG moments, rather than independent ones for $\epem$ and DIS
collisions, in order to extract a more precise value of $\alphas$.

\section{Theoretical 
framework}
\label{sec:theory}

The parton-to-hadron FF, $D_{\rm i\to h}(z,\rm Q)$, encodes the probability that parton 
$i$ fragments into a hadron $h$ carrying a fraction $z$ of the parent parton's momentum. 
FFs for gluon and different flavors of quark-initiated jets can be computed 
from DGLAP evolution equations \cite{Gribov:1972ri,Altarelli:1977zs,Dokshitzer:1977sg} 
in perturbation theory. As for the Schr\"odinger equation in quantum mechanics, the 
system of equations can be written as an evolution {\em Hamiltonian} which mixes gluon 
and (anti)quark states expressed in terms of DGLAP splitting functions for
the branchings $g\to gg$, $q(\bar{q})\to gq(\bar{q})$ and $g\to q\bar{q}$, where $g$, 
$q$ and $\bar{q}$ label a gluon, a quark and an antiquark respectively. Analytical 
solutions can be obtained by using a  Mellin transform over the convolution product 
of the regularized splitting functions and the FFs with respect to $\xi$. 
For soft partons, the shift in $\xi$ is reabsorbed into the exponential such that
the energy radiated by the parton $\omega$ can be replaced by
$\omega\to\Omega=\omega+\partial/\partial \rm Y$ (where Y is related to the (log) 
of the energy of the initial parton), resulting in a final expansion in half-powers of $\alphas$. 
In order to incorporate ${\cal O}(\alphas^{3/2})$ contributions, going beyond the
${\cal O}(\alphas)$ terms obtained in older approaches, the matrix elements of the 
evolution {\em Hamiltonian} should be expanded up to terms $\propto\Omega$, 
followed by its diagonalisation, which results into two eigenvalues $P_{\pm\pm}(\Omega)$ 
in the new ${\cal D}^{\pm}(\Omega,\rm Q)$ basis. This procedure leads to the 
following equation for the eigenvector ${\cal D}^+$~\cite{Perez-Ramos:2013eba}, 
\vbox{
\begin{eqnarray}
&&\hspace{-1.2cm}
\left(\omega+\frac{\partial}{\partial \rm Y}\right)\frac{\partial}{\partial \rm Y}{\cal D}^+(\omega,{\rm Y},\lambda)
=\left[1-\frac{a_1}{4N_c}\left(\omega+\frac{\partial}{\partial \rm Y}\right)\right.\cr
&&\left.+a_2\left(\omega
+\frac{\partial}{\partial \rm Y}\right)^2\right]
4N_c\frac{\alphas}{2\pi}{\cal D}^+(\omega,\rm Y,\lambda)
\label{eq:gluonDg}
\end{eqnarray}
}
which is the one that provides a Gaussian-like shape for the distribution, while 
${\cal D}^-$ vanishes asymptotically. As explained 
in~\cite{Perez-Ramos:2013eba}, $a_1$ and $a_2$ are hard constants depending on the 
number of active flavors $N_f$ and on the $C_F$ and $N_c$ Casimirs of the fundamental and adjoint
representation of the SU(3) color group respectively; and $\lambda= \ln(\rm Q_0/\lqcd)$ is the
hadronization parameter at which the shower stops. The terms $\propto a_1$ and $a_2$ provide respectively NLL
and NNLL corrections. 
Equation~(\ref{eq:gluonDg}) 
is solved by using the two-loop expression of $\alphas$:
\begin{equation}\label{eq:twoloop}
\hspace{-0.16cm}\alphas(q^2)=\frac{4\pi}{\beta_0\ln q^2}\left[1-\frac{2\beta_1}{\beta_0^2}\frac{\ln\ln q^2}
{\ln q^2}\right],  \mbox{ with }\, q^2=\frac{\rm Q^2}{\lqcd^2},
\end{equation}
and $\beta_{0,1}$ are the first two coefficients of the perturbative expansion of the
$\beta$-function through the renormalisation group equation~\cite{Caswell:1974gg}.\\

The solution of Eq.~(\ref{eq:gluonDg}) can be written in the compact form:
\begin{equation}\label{eq:Dgomega}
{\cal D}^+(\omega,\rm Y,\lambda)=E_+(\omega,\alphas(\rm Y+\lambda)){\cal D}^+(\omega,\lambda),
\end{equation}
with the evolution {\em Hamiltonian} rewritten in terms of the anomalous dimension 
$\gamma(\omega,\alphas)$ as:
\begin{equation}\label{eq:Egg}
E_+(\omega,\alphas(\rm Y+\lambda))=\exp\left[\int_{0}^{\rm Y}dy\,\gamma(\omega, \alphas(y+\lambda))
\right],
\end{equation}
with $y=\rm Y-\xi$, which describes the parton jet evolution from its initial 
virtuality Q to the lowest possible energy scale Q$_0$, at which the parton-to-hadron
transition occurs. Inserting Eq.~(\ref{eq:Egg}) into (\ref{eq:gluonDg}),
the resulting equation for $\gamma(\omega,\alphas)$ can be solved iteratively.
Its solution at NLO*+NNLL accuracy reads:
\begin{eqnarray}\label{eq:nmllagamma}
\hspace{-0.6cm}\gamma_\omega^{_{\rm NLO^*+NNLL}}\!\!&\!\!=\!\!&\!\!\gamma_\omega^{_{\rm MLLA}}+\frac{\gamma_0^4}{16N_c^2}
\left\{a_1^2\frac{\gamma_0^2}{(\omega^2+4\gamma_0^2)^{3/2}}\right.\cr
\hspace{-0.5cm}\!\!&\!\!+\!\!&\!\!\left.\frac{a_1\beta_0}{2}\left(\frac1{\sqrt{\omega^2+4\gamma_0^2}}-\frac{\omega^3}{(\omega^2+4\gamma_0^2)^2}\right)
\right.\cr
\hspace{-0.6cm}\!\!&\!\!+\!\!&\!\!\left.\beta_0^2\left(\frac{2\gamma_0^2}
{(\omega^2+4\gamma_0^2)^{3/2}}-\frac{5\gamma_0^4}{(\omega^2+4\gamma_0^2)^{5/2}}\right)\right.\cr
\hspace{-0.6cm}\!\!&\!\!-\!\!&\!\!\left.4N_c\frac{\beta_1}{\beta_0}\frac{\ln2(\rm Y+\lambda)}{\sqrt{\omega^2+4\gamma_0^2}}
\right\}\cr
\hspace{-0.6cm}\!\!&\!\!+\!\!&\!\!\frac{1}{4}a_2\gamma_0^2\left[\frac{\omega}{(\omega^2+4\gamma_0^2)^{1/4}}
+(\omega^2+4\gamma_0^2)^{1/4}\right]^2,
\end{eqnarray}
where $\gamma_\omega^{_{\rm MLLA}}$ is the MLLA anomalous dimension computed first
by Fong \& Webber~\cite{Fong:1990nt} and $\gamma_0\sim\sqrt{\alphas}$ is the anomalous 
dimension obtained in the double logarithmic approximation (DLA)~\cite{Fadin:1983aw}. 
The MLLA anomalous dimension resums double soft-collinear leading logarithms (DLA, of order 
${\cal O}(\sqrt{\alphas})$) and  hard-collinear next-to-leading logs (single logarithms, of order ${\cal O}(\alphas)$) 
which partially restore energy conservation and account for LO running coupling effects. 
Terms proportional to $\beta_1$ and $a_2$  are corrections computed for the first time in our 
NMLLA+NLO* framework. In the expression proportional to $\gamma_0^4$, the terms 
$\propto a_1^2,a_1\beta_0, \beta_0^2$ mix NNLL contributions obtained iteratively, the 
$\propto a_2$ term resums next-to-next-to-leading logarithms  which 
improve energy conservation at each branching vertex of the parton shower 
and, finally, the $\propto\beta_1$ term accounts for NLO running coupling effects. 
The overall new correction is ${\cal O}(\alphas^{3/2})$ as can be checked 
from the power counting in Eq.~(\ref{eq:nmllagamma}), knowing that 
$\omega\sim{\cal O}(\alphas^{-1/2})$.\\

\begin{figure*}[!htbp]
\begin{center}
\includegraphics[width=7.5cm,clip]{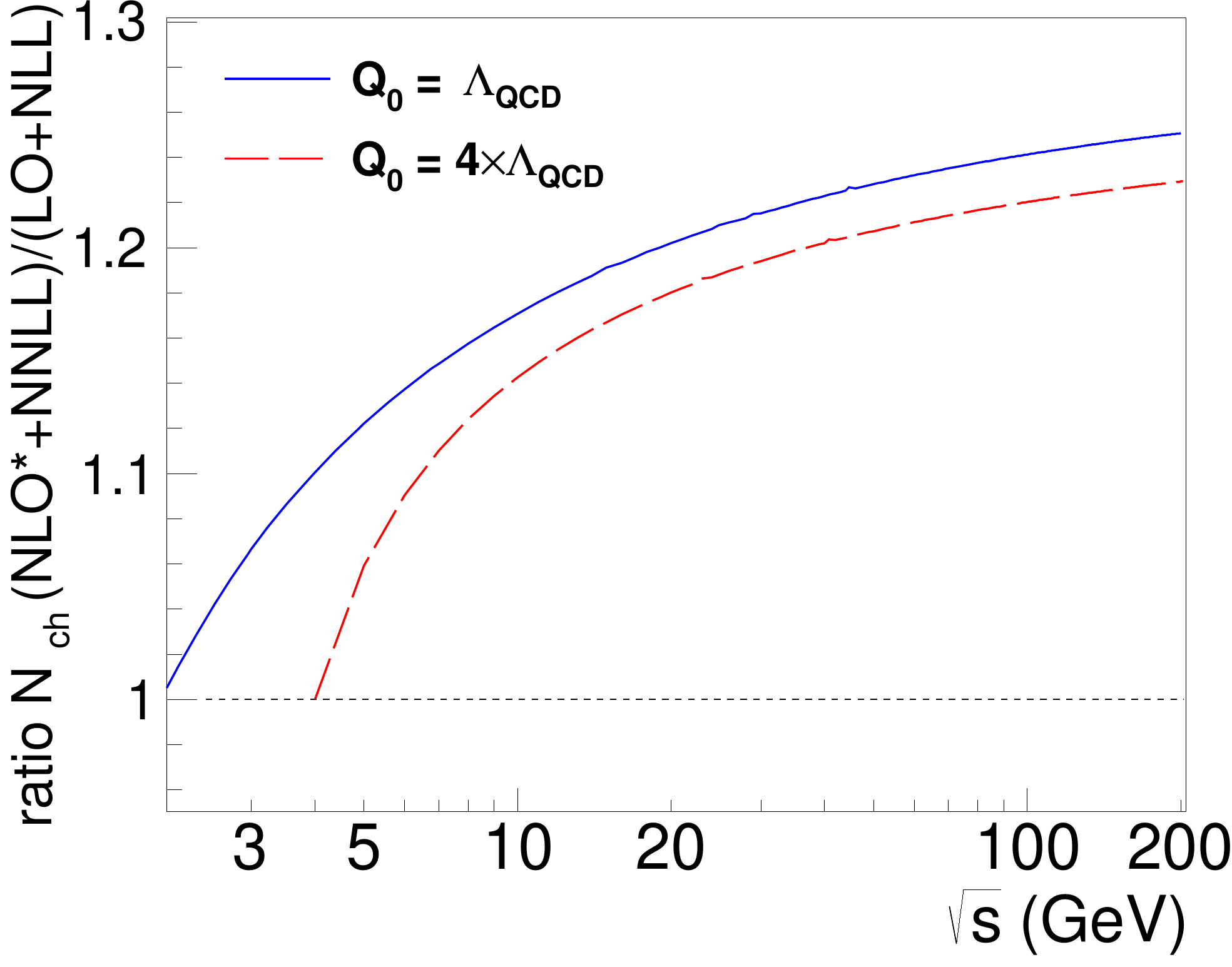}\hspace{2mm}
\includegraphics[width=7.5cm,clip]{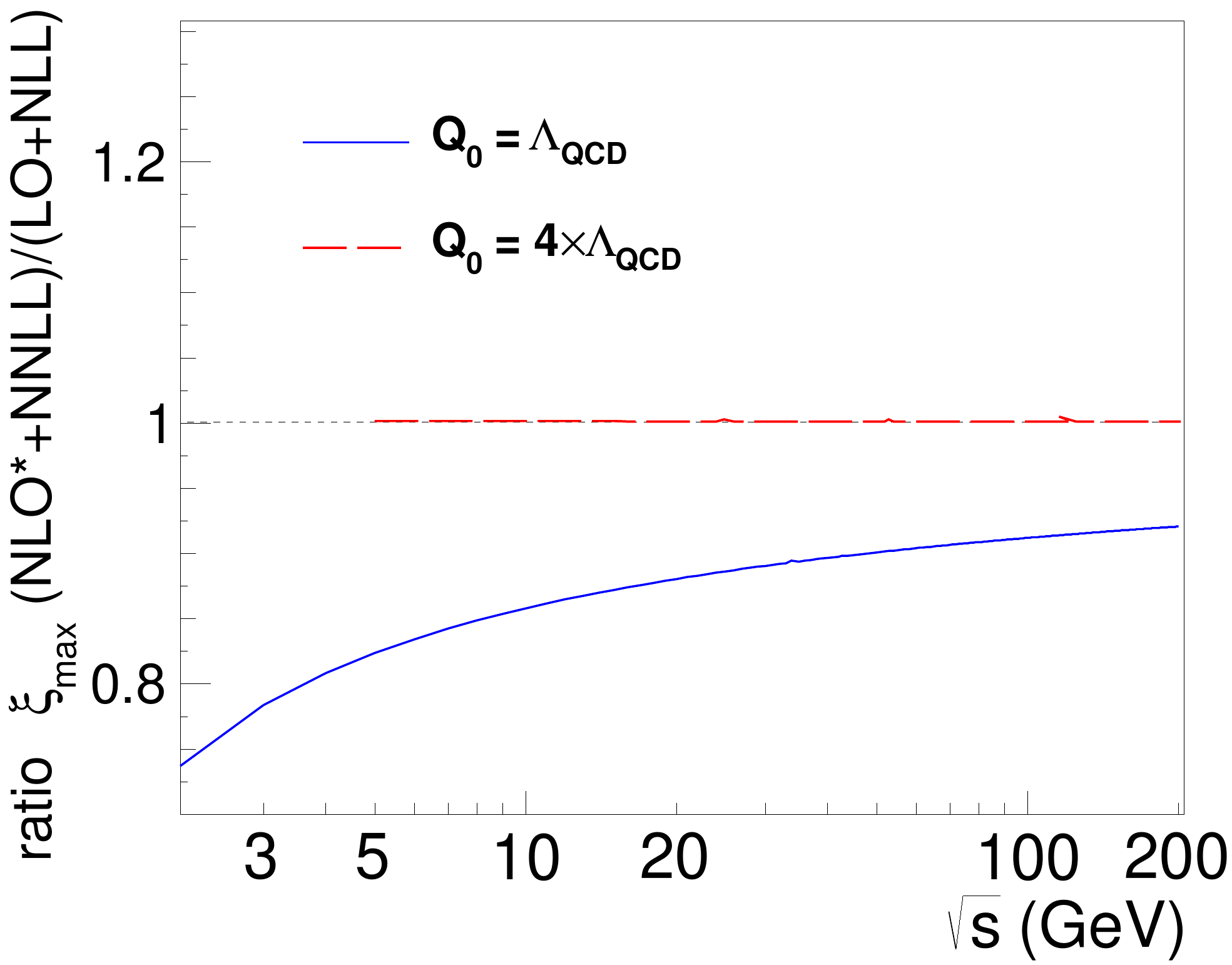}
\includegraphics[width=7.5cm,clip]{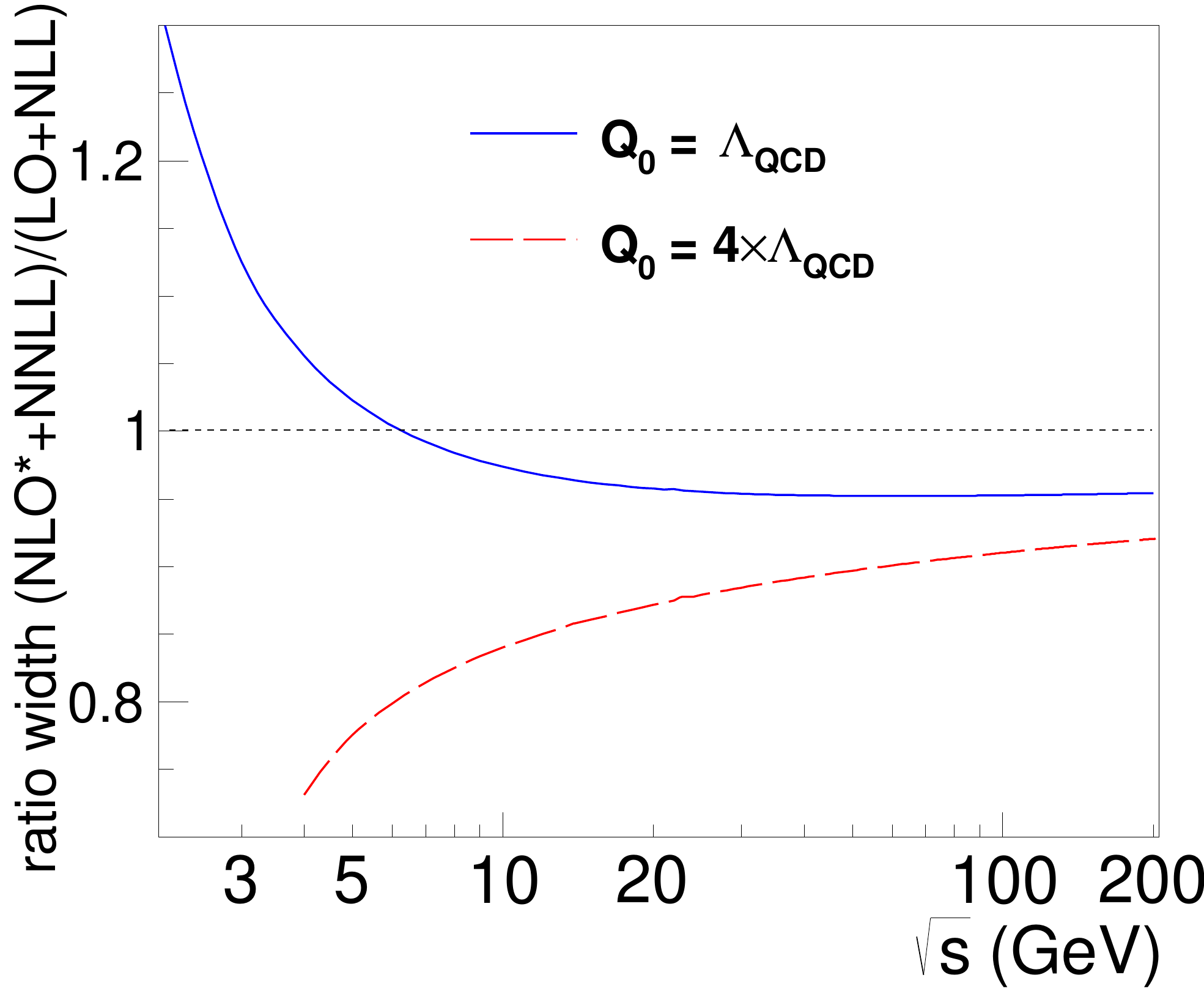}\hspace{2mm}
\includegraphics[width=7.5cm,clip]{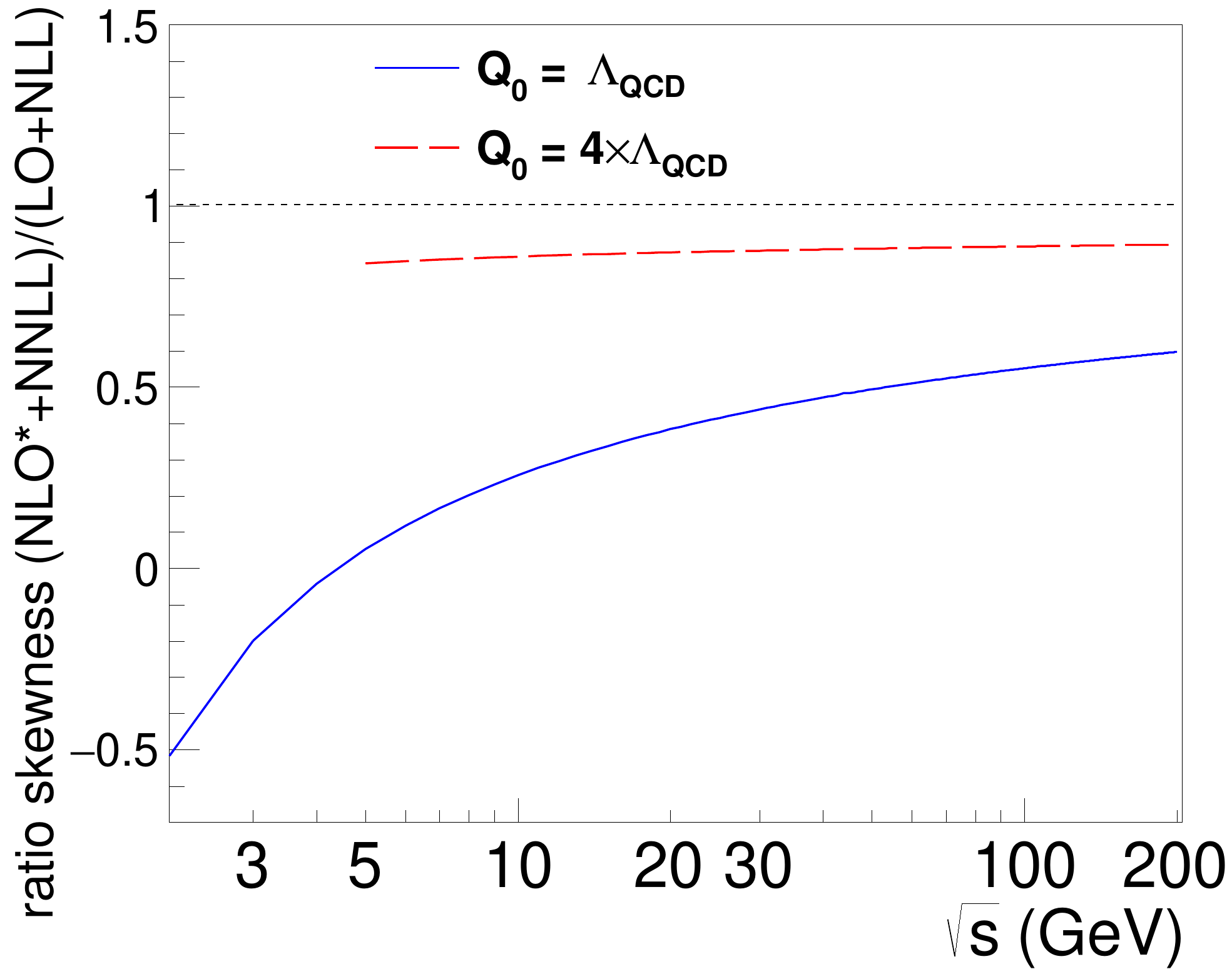}
\caption[]{Ratios of the theoretical predictions for the energy evolution of the moments of the
  parton-to-hadron FFs: multiplicity (top-left), peak (top-right), width (bottom-left) and skewness
  (bottom-right) at NLO*+NNLL over LO+NLL for two values of the shower cutoff Q$_0 = \lqcd$
  (solid curve) and Q$_0 = 4\cdot \lqcd \approx$~1~GeV (dashed curve).
}
\label{fig:ratios}
\end{center}
\end{figure*}

Replacing the Mellin transform of Eq.~(\ref{eq:DG}) into Eq.~(\ref{eq:Dgomega}),
the DG moments can be obtained at NLO*+NNLL accuracy from the anomalous 
dimension~(\ref{eq:nmllagamma}) by comparing both sides of the 
resulting equation, via
\begin{equation}\label{eq:moments}
K_{n\geq0}(\rm Y,\lambda)=\int_0^{\rm Y}dy\left(-\frac{\partial}
{\partial\omega}\right)^n\gamma_\omega(\alphas(y+\lambda))\bigg|_{\omega=0},
\end{equation}
with
\begin{equation}\label{eq:K1234}
{\cal N}= K_0,\quad \bar\xi= K_1,\;
\sigma=\sqrt{K_2},\; s=\frac{K_3}{\sigma^3},\; k=\frac{K_4}{\sigma^4}.
\end{equation}
The full expressions of the energy evolution of the moments (\ref{eq:K1234}) have been derived as a function of 
Y~=~$\ln(E/\lqcd)$ (for an initial parton energy $E$) 
and $\lambda$ for $N_f=3,4,5$ 
in~\cite{Perez-Ramos:2013eba}. The resulting formulae for the energy evolution of the moments
depend on $\lqcd$ as {\it single} free parameter. Particularly simple expressions are obtained in the
limiting-spectrum case ($\lambda=0$, i.e. for Q$_0 = \lqcd$) 
motivated by the ``local parton hadron duality'' hypothesis for infrared-safe observables 
which states that the distribution of partons in jets are simply renormalized in the hadronization process
without changing their HBP shape. 
In Fig.~\ref{fig:ratios} we display the ratios (NLO*+NNLL)/(LO+NLL) of the moments so as to shed 
more light on the size of the new corrections of Eq.~(\ref{eq:nmllagamma}) compared to 
older LO+NLL results. We plot the ratios for the limiting-spectrum case 
and for a scale $\lambda$~=~1.4 (i.e. for Q$_{0} = 4\,\lqcd\approx$~1~GeV). Various observations 
are worth pointing out. First, the NLO*+NNLL corrections are quite sizable for all DG moments in 
the limiting spectrum but smaller for larger values of the shower energy cutoff. 
This is not surprising since, as $\lambda$ increases, the convergence of the 
perturbative expansion is improved, higher order corrections decrease much faster 
and the (NLO*+NNLL)/(LO+NLL) ratios approach unity in each case (solid curves), as observed.
Second, focusing in the simplest limiting-spectrum case, we see that the impact of higher-order corrections is
different in size and in sign for the different FF moments. On the one hand, the predictions for the multiplicity
are larger by up to a 25\% in the NLO*+NNLL framework compared to the LO+NLL predictions. On the other, the DG
peak and skewness are smaller at NLO*+NNLL by about 10--20\% and more than 50\% respectively, whereas the DG
width evolution varies by $\pm$(10--20)\% depending on the jet energy. For a fixed jet energy, higher hadron 
multiplicity, DG peak and width translate into comparatively lower values of the associated $\lqcd$, whereas
higher values of the DG skewness reflect larger $\lqcd$. Those results highlight the importance of properly
accounting for higher-order contributions in any combined analysis of the energy evolution of the FF moments.
Work is in progress to include the full-NLO (and beyond) corrections of the evolution of the FF moments~\cite{DdE_RP}.

%
\section{Extraction of $\boldsymbol{\alphas}$ }

The procedure of extraction of $\alphas$ from the experimental data consists in a two-step process.
First, we collect all the existing parton-to-hadron FFs measured so far and fit them to the DG
expression (\ref{eq:DG}) in order to obtain four FF moments at each jet energy. 
Finite hadron-mass effects in the DG fit have been accounted for through a rescaling of the theoretical
(massless) parton momenta with an effective mass\footnote{Since the measured FFs are for
massive hadrons and the calculations assume massless partons/hadrons (for which $\xi_{\rm p}=\xi_{\rm E}$), 
the expression (\ref{eq:DG}) for $\xi$ needs to be modified by introducing an effective mass,
$E=\sqrt{p^2+\meff^2}$, plus the corresponding Jacobian determinant correction.} 
$\meff$ as discussed in Ref.~\cite{Perez-Ramos:2013eba}.
Next, we carry out a combined fit of the four moments as a function of the original parton energy (which in
the case of $\epem$ collisions corresponds to half the centre of mass energy $\sqrts/2$ and, for DIS, to the invariant 
four-momentum transfer ${\cal Q}$) to our NMLLA+NLO$^*$ predictions leaving $\lqcd$ as a free parameter in the
fit. From the extracted $\lqcd$, we obtain the value of the QCD coupling at the Z pole, $\alphasmZ$, using 
the two-loop running Eq.~(\ref{eq:twoloop}) for $N_f=5$ quark flavours.

\begin{figure*}[htpb!]
\centering
\includegraphics[width=0.90\linewidth,height=11.9cm]{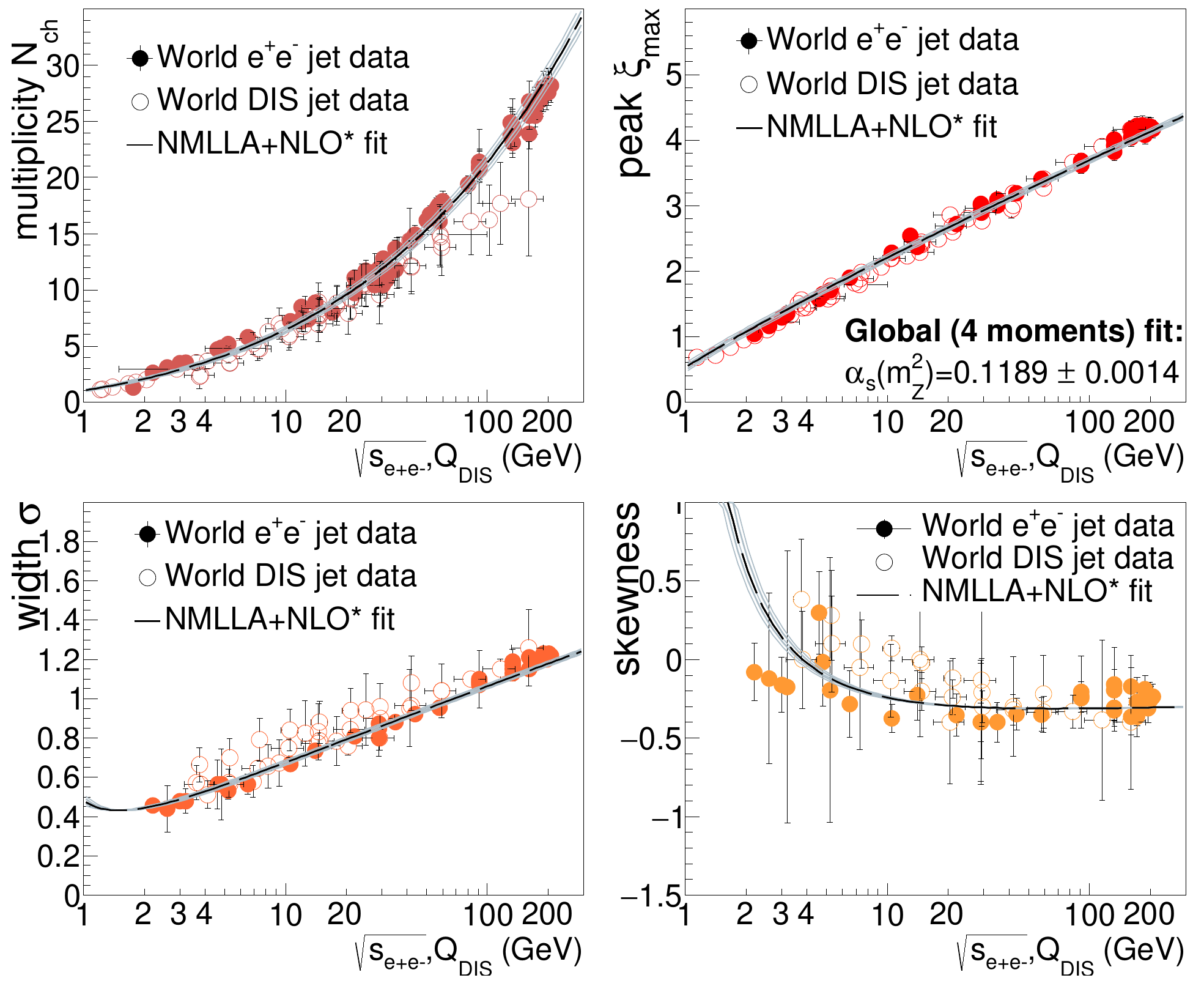}
\caption[]{Global NLO*+NNLL fit of the energy evolution of the moments (charged hadron multiplicity, peak,
  width and skewness) of the jet FFs measured in $\epem$ collisions at 
$\sqrts_{_{\rm e+e-}}\approx$~2--200~GeV (solid circles) and $e^\pm,\nu$-p collisions 
at ${\cal Q}_{_{\rm DIS}}\approx$~4--180~GeV (open circles). 
The extracted value of $\alphasmZ$ is quoted in top-right panel.} 
\label{fig:3}
\end{figure*}

\subsection{Data sets and fits}

Our analysis includes 37 measurements of FFs in $\epem$ collisions covering the range of c.m. 
energies $\sqrts\approx$~2--200~GeV from the following experiments:
BES at $\sqrts$~=~2--5~GeV~\cite{BES:2003xt}; BaBar\footnote{The individual distributions measured for 
prompt pions, kaons and (anti)protons have been added into a single charged-particle distribution.} 
at $\sqrts$~=~10.54~GeV~\cite{Lees:2013rqd};
MARK-II at $\sqrts$~=~5.2, 6.5 and 29~GeV~\cite{Patrick:1982pp};
TASSO at $\sqrts$~=~14--44~GeV~\cite{Braunschweig:1988qm,Braunschweig:1990yd}; 
TPC/Two-Gamma  at $\sqrts$~=~29~GeV~\cite{Aihara:1988su}; 
HRS at $\sqrts$~=~29~GeV~\cite{Bender:1984fp};
TOPAZ at $\sqrts$~=~58~GeV~\cite{Itoh:1994kb}; 
ALEPH~\cite{Barate:1996fi}, L3~\cite{Adeva:1991it} and OPAL~\cite{Akrawy:1990ha,Ackerstaff:1998hz} at $\sqrts$~=~91.2~GeV; 
ALEPH~\cite{Buskulic:1996tt,Heister:2003aj}, DELPHI~\cite{Abreu:1996mk} and OPAL~\cite{Alexander:1996kh} 
at $\sqrts$~=~133~GeV; and ALEPH~\cite{Heister:2003aj} and
OPAL~\cite{Ackerstaff:1997kk,Abbiendi:1999sx,Abbiendi:2002mj} in the range $\sqrts$~=~161--202~GeV.
The analysis presented here extends our previous study~\cite{Perez-Ramos:2013eba} with 5 new
datasets~\cite{Lees:2013rqd,Patrick:1982pp,Bender:1984fp}.
The total number of FF points is about 1000 (Fig.~\ref{fig:DGfits} left) and the systematic and statistical
uncertainties of each single-hadron spectrum have been added in quadrature. 
Beyond the results of our DG fits, we add also other FF moments which have been directly measured in
$\epem$ collisions (not already included in the FF fits above) such as:  
41 $N_{\rm ch}$ values in the range $\sqrts$~=~12--161~GeV compiled in~\cite{Lafferty:1995jt}, 
3 $N_{\rm ch}$ and $\xi_{\rm max}$ values measured by the JADE collaboration at
$\sqrts$~=~12, 30, 35~GeV~\cite{Bartel:1983qp}, plus the average charged multiplicity measured 
in the world data~\cite{PDG} of hadronic decays of the Z boson ($\mean{\rm N_{\rm ch}}$~=~20.76~$\pm$~0.16), 
the W boson ($\mean{\rm N_{\rm ch}}$~=~19.39~$\pm$~0.08), and the $\tau$ lepton 
($\mean{\rm N_{\rm ch}}$~=~1.314~$\pm$~0.002) 
corresponding to $\sqrts = m_{_{\rm Z,W},\tau}\approx$~91.2, 80.4, 1.77~GeV respectively. 
The total final number of FF moments extracted from the world $\epem$ jet data amounts to 200. 
In the case of DIS collisions, our analysis fits first the single-hadron distributions (amounting to about 
250 individual data points, see Fig.~\ref{fig:DGfits} right) measured by ZEUS~\cite{ZEUS} in the current
hemisphere of the Breit (or ``brick wall'') frame where the incoming quark scatters off the photon and returns
along the same axis. In addition, we include also into our global fit the 55 direct measurements of FF moments
(mostly multiplicity, peak, and width) from H1~\cite{H1old} and ZEUS~\cite{ZEUSold} experiments in $e^\pm$-p at
HERA, and NOMAD ($\nu$-N scattering)~\cite{Altegoer:1998py}, covering the range of four-momentum transfers
${\cal Q}\approx$~1--180~GeV. The total final number of FF moments extracted from the DIS jet data is 160 and, thus, the
combined global fit of $\epem$ and e$^\pm$,$\nu$-p moments amounts to 360 data points. 

\begin{figure*}[!htbp]
\begin{center}
\includegraphics[width=7.5cm,height=5.5cm]{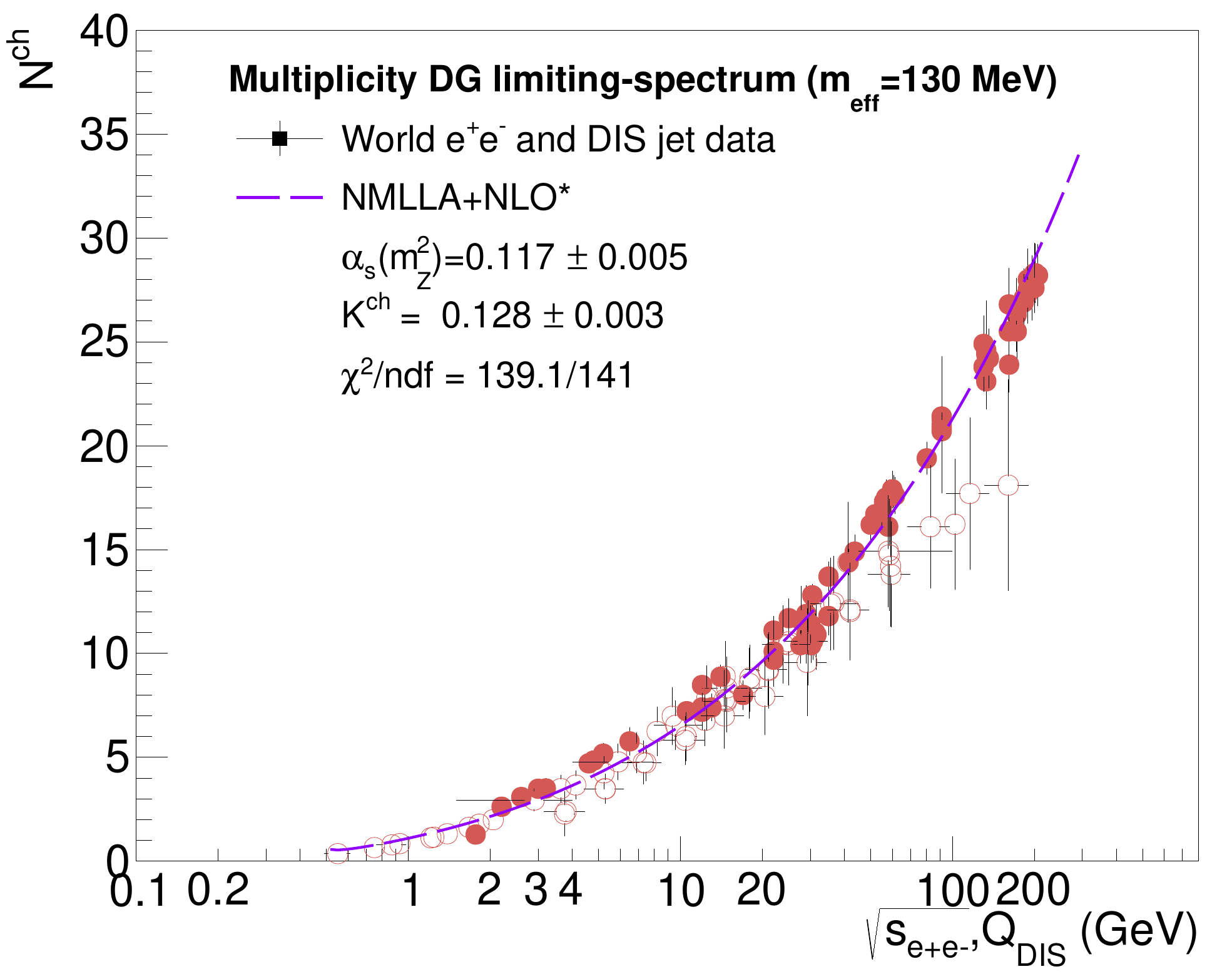}\hspace{2mm}
\includegraphics[width=7.5cm,height=5.5cm]{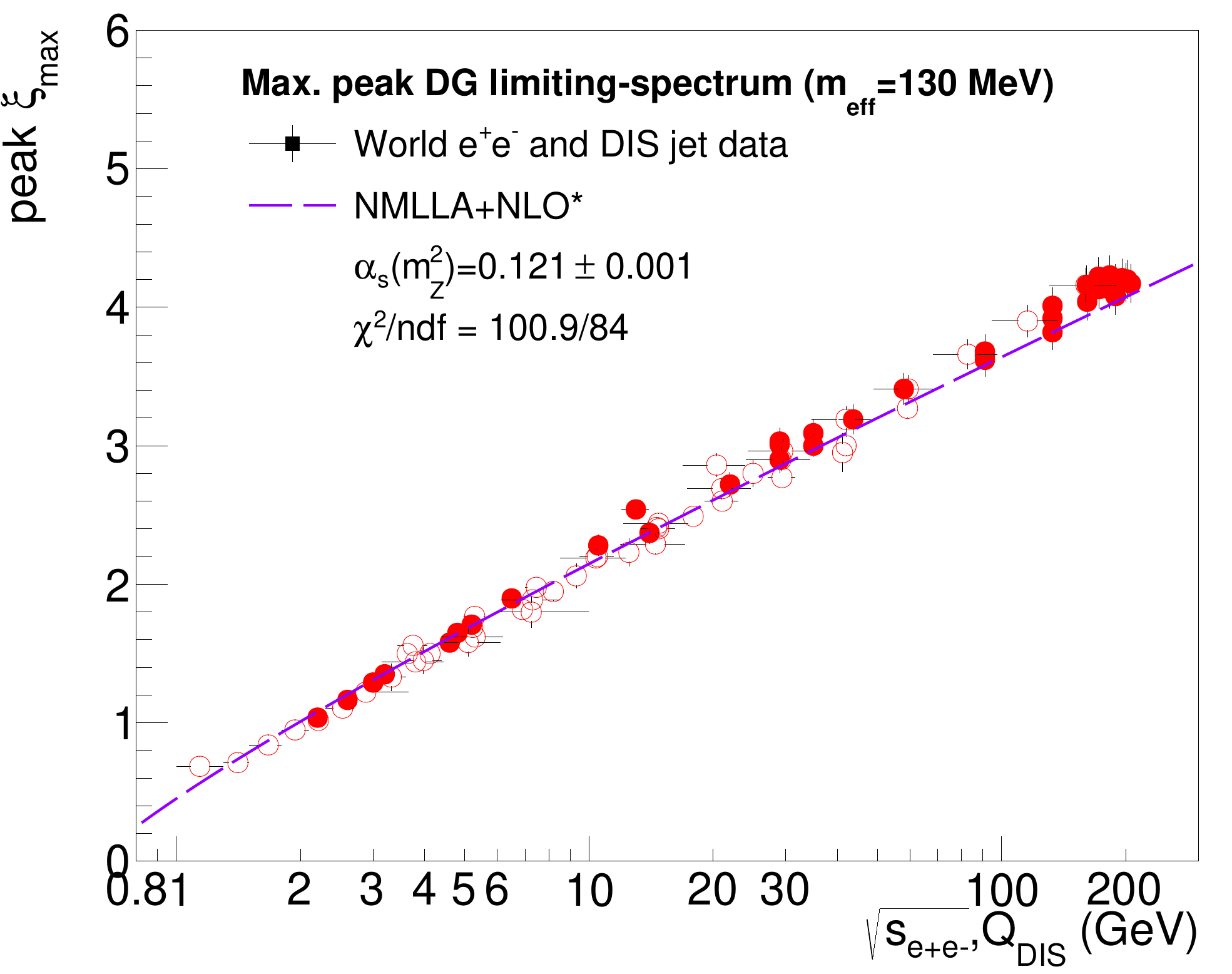}
\includegraphics[width=7.5cm,height=5.5cm]{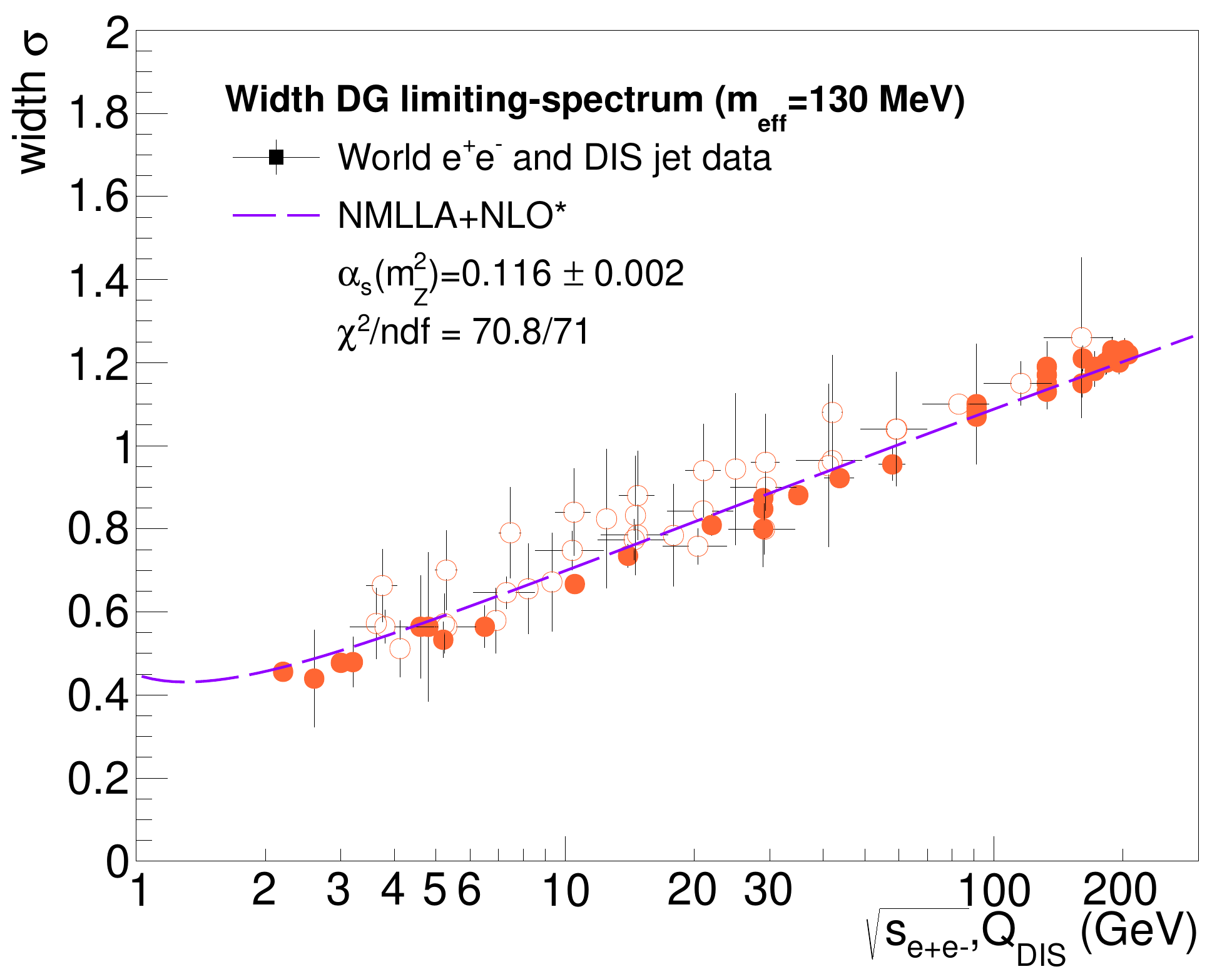}\hspace{2mm}
\includegraphics[width=7.5cm,height=5.5cm]{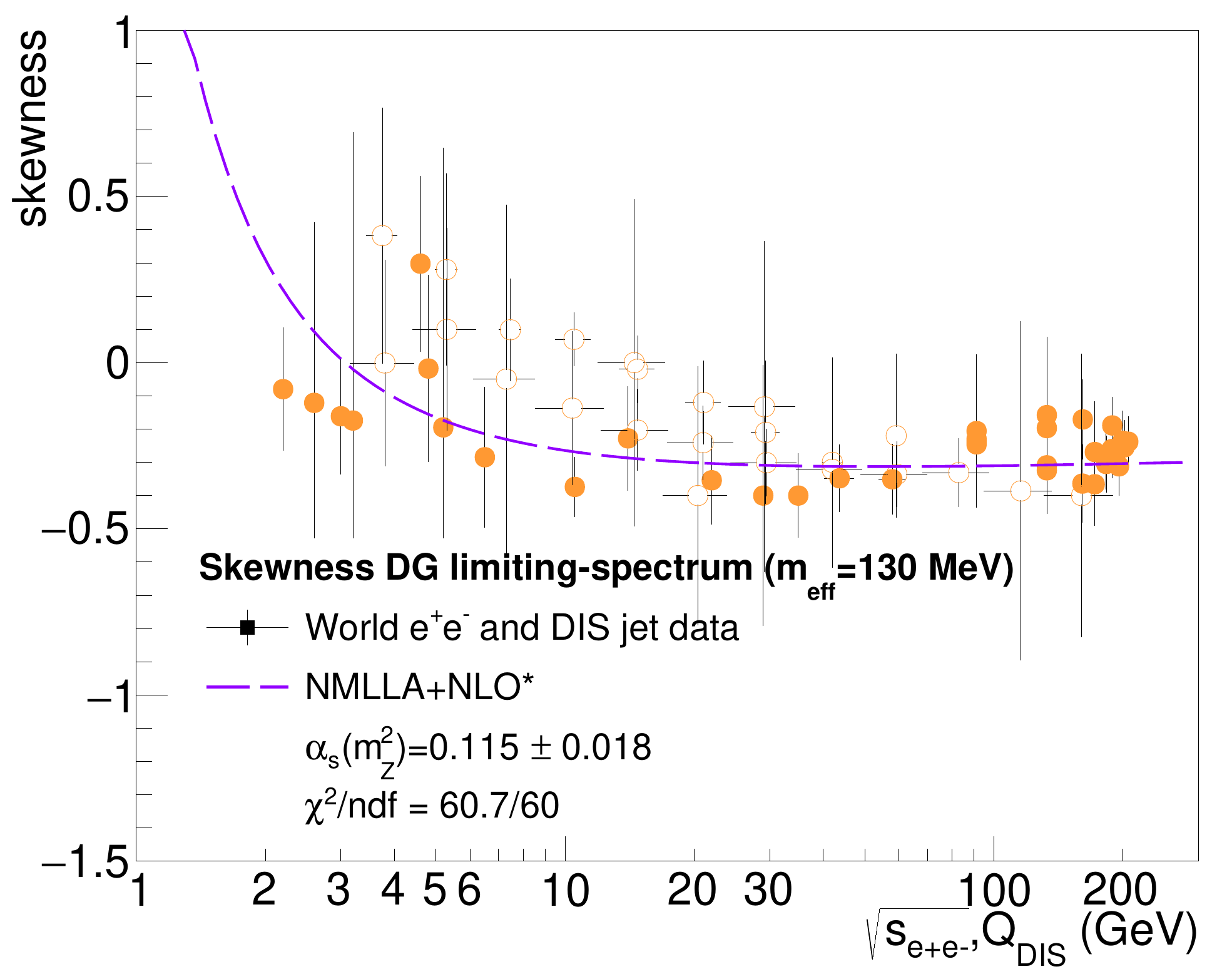}
\caption[]{Individual NLO*+NNLL fits of the energy dependence of the moments of the parton-to-hadron FFs
  measured in $\epem$ and DIS collisions: multiplicity (top-left), peak (top-right), width (bottom-left) and
  skewness (bottom-right); after $\chi^2$-reweighting of some DIS data (see text). The extracted values of
  $\alphasmZ$ (and associated goodness-of-fit $\chi^2$/ndof) are quoted for each fit.
}
\label{fig:IndividualFits}
\end{center}
\end{figure*}

The combined fit of the energy-dependencies of the multiplicity, peak, width and skewness of the experimental
FFs to our NLO*+NNLL predictions for the evolution of the limiting spectrum FF moments, leaving
$\lqcd$ as free parameter, is shown in Fig.~\ref{fig:3}. The fit includes also corrections for each moment 
to account for the increasing number of flavours, $N_f$~=~3,4,5, at the corresponding heavy-quark production
thresholds: $E_{\rm jet}>$~m$_{\rm charm,\,bottom}\approx$~1.3,\,4.2~GeV. The global fit, obtained using 
the {\sc  minuit2} package (with the {\tt MIGRAD} minimizer, although alternative algorithms give identical
results)~\cite{James:1994vla} implemented in {\tt ROOT}, yields a QCD coupling strength at the Z mass pole of
$\alphasmZ$~=~0.1189~$\pm$~0.0014. The values obtained from separate fits of our previous (smaller) sets of
$\epem$ and DIS data are: $\alphasmZ$~=~0.1195~$\pm$~0.0022~\cite{Perez-Ramos:2013eba} and
0.119~$\pm$~0.010~\cite{d'Enterria:2014yya}, respectively. 
In the range of experimental jet energies considered, $E_{\rm jet}\approx$~1--100~GeV, 
the average scale at which $\alphas$ is effectively evaluated in our approach is given by 
the geometric mean between the energy of the original parton and that at the end of the 
shower evolution, 
i.e. $\mean{\rm Q}=\sqrt{E_{\rm jet} \cdot \rm Q_{0}}\approx$~0.6--2~GeV in the limiting spectrum case.

\subsection{Systematic uncertainties} 

Our extracted value of $\alphasmZ$~=~0.1189~$\pm$~0.0014 has a relative uncertainty of about 1.2\% 
(without theoretical scale uncertainties which are discussed below), which is a very competitive value
compared to other existing $\alphas$ determinations~\cite{PDG}. 
The $\alphasmZ$ uncertainties have been obtained through the ``$\chi^2$ averaging'' method~\cite{PDG} as
explained in~\cite{d'Enterria:2014bsa}: We fit first the energy-dependence for each individual moment to its
corresponding theoretical prediction (Fig.~\ref{fig:IndividualFits}) and if the goodness-of-fit $\chi^2$ is
larger than the number of degrees of freedom (ndof), then the data points of the corresponding moment are
enlarged by a common factor such that $\chi^2$/ndof equals unity. As a matter of fact, such an error
enlargement has been applied only to two DIS moments: N$_{\rm ch}$ (which requires an overall $+$20\% increase
in its uncertainties) and width $\sigma$ ($+$10\% increase), which show a larger scatter in their central values
compared to the $\epem$ measurements. 
In particular, as can be seen from the top-left panel of Fig.~\ref{fig:IndividualFits}, the  hadron
multiplicities measured in DIS jets 
are systematically $\sim$20\% smaller (especially at the highest energies) than those measured in $\epem$
collisions~\cite{Perez-Ramos:2013eba}, a  fact pointing maybe to limitations in the FF measurement only in
half (current Breit) $e^\pm$-p hemisphere.
The $\chi^2$-averaging method takes into account in a well defined manner possible correlations among the
evolutions of the four FF moments, as well as any missing extra systematic uncertainties. Our detailed
assessment of the experimental uncertainties~\cite{d'Enterria:2014bsa} indicates that such an error assignment 
covers perfectly well the range of $\alphas$ variations induced by the existing
sources of uncertainty: 
(i) finite hadron-mass effects corrected through an $\meff$ factor introduced in
Eq.~(\ref{eq:DG}), and 
(ii) the use of data-sets with slightly different definitions of final charged hadrons
(including, or not, a fraction of secondary hadrons from weak K$_s^0$ and $\Lambda$ decays).
Both effects have been estimated in~\cite{d'Enterria:2014bsa} by scrutinizing the 
high-precision prompt and inclusive hadron FFs measured by the BaBar experiment~\cite{Lees:2013rqd}.
We find that our FF analysis is robust with respect to hadronization and other experimental uncertainties.
Indeed,
\begin{description}
\item (i) The fit uncertainties for all extracted moments cover well the variations due to suitable
choices of effective hadron masses in the range $\meff\!\!$~=~0--0.2~GeV. The fits shown in
Fig.~\ref{fig:DGfits} have been obtained for effective masses that result in the best goodness-of-fit
($\meff\,$=~130~MeV for $\epem$, and 110~MeV for DIS), but conservative $\pm$50\% variations from these
default values yield consistent $\alphas$ extractions.
\item (ii) The DG moments obtained from the FFs for prompt (primary) and inclusive (primary+decay)
charged hadrons are all consistent within their associated uncertainties except, as expected, for the total
multiplicity N$_{\rm ch}$ which is a factor of (9$\pm$1)\% smaller for the primary-hadrons FFs. Also, the
analysis shows that older DIS extractions of the FF width using a simple Gaussian function overestimate
their values by 20\% compared to the more realistic DG fits\footnote{Note that the original FFs are not
available in some of the oldest DIS measurements~\cite{H1old} and their measured Gaussian widths
have been included into our global energy-evolution fit with such a correction factor applied.}.
\end{description}

\begin{figure}[htpb!]
\includegraphics[width=0.99\linewidth]{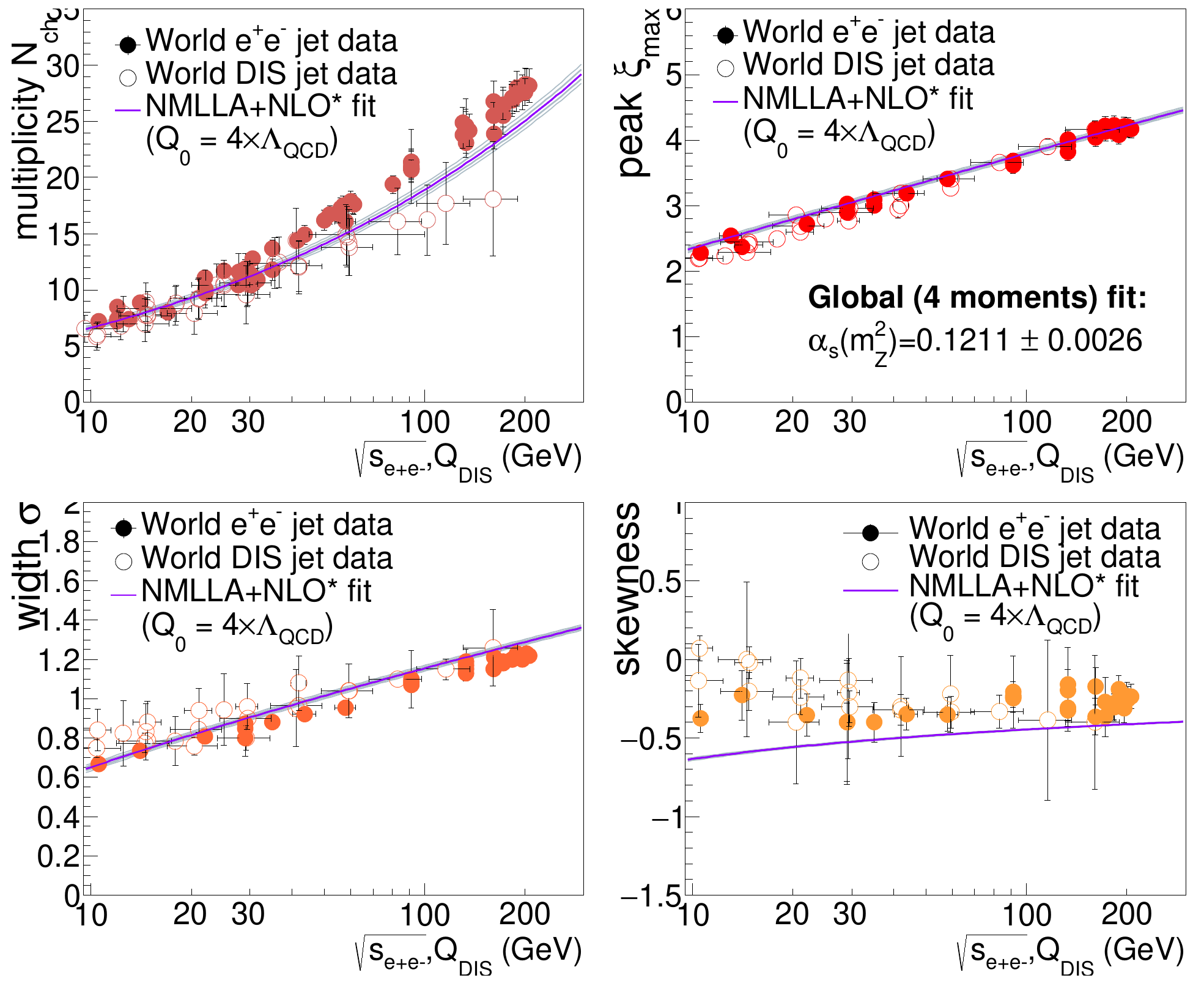}
\caption[]{Energy evolution of the moments (charged hadron multiplicity, peak, width and skewness) 
of the jet FFs measured in $\epem$ and DIS collisions at $\sqrts\approx$~10--200~GeV, fitted to the NLO*+NNLL
predictions evaluated at a scale $\lambda$~=~1.4 (i.e. Q$_{0} = 4\cdot\lqcd\approx$~1~GeV).}
\label{fig:4}
\end{figure}

The last source of systematic uncertainty is of purely theoretical nature and it is associated with missing
fixed-order terms in our truncation of the $\alphas$ expansion at approximate NLO accuracy. Although our $\alphas$
determination relies on a theoretically framework that resums soft and collinear logs down to Q$_{0} = \lqcd$,
and thus it is much more robust with respect to hadronization corrections than other methods which treat
such effects as extra non-perturbative uncertainties, the state-of-the-art $\alphas$ determinations have
one-level of higher (NNLO) accuracy~\cite{PDG}. In order to estimate the size of theoretical scale 
uncertainties associated with missing higher-order terms, we have redone the analysis for a different energy
scale $\lambda$ at which the parton evolution is stopped. Using $\lambda$~=~1.4 (i.e. 
Q$_{0} = 4\cdot\lqcd\approx$~1~GeV) and limiting the energy evolution fits to jets energies in the range
$E$~=~10--200~GeV, so as to leave enough room for parton evolution avoiding data with energies too close to the
shower cutoff at 1~GeV, we obtain $\alphasmZ$~=~0.1211$\pm$0.0026 (Fig.~\ref{fig:4}). We see that the
data-theory agreement is good for all FF moments except for the skewness which shows a better fit with the
default limiting-spectrum ansatz. 
The QCD coupling value obtained stopping the parton evolution of the FFs at 1~GeV is consistent with that
determined in the limiting spectrum case, $\alphasmZ$~=~0.1189~$\pm$~0.0014, although larger by +0.0022. We
conservatively assign this difference as a (positive) source of systematic error associated with the scale
uncertainty of our calculations. Adding in quadrature this asymmetric error to the previously determined
$\pm$0.0014 systematics uncertainty, results in the final quantitative result of our study: 
$\alphasmZ$~=~0.1189$^{+0.0025}_{-0.0014}$. 

\begin{figure}[htpb!]
\centerline{
\includegraphics[width=0.99\linewidth]{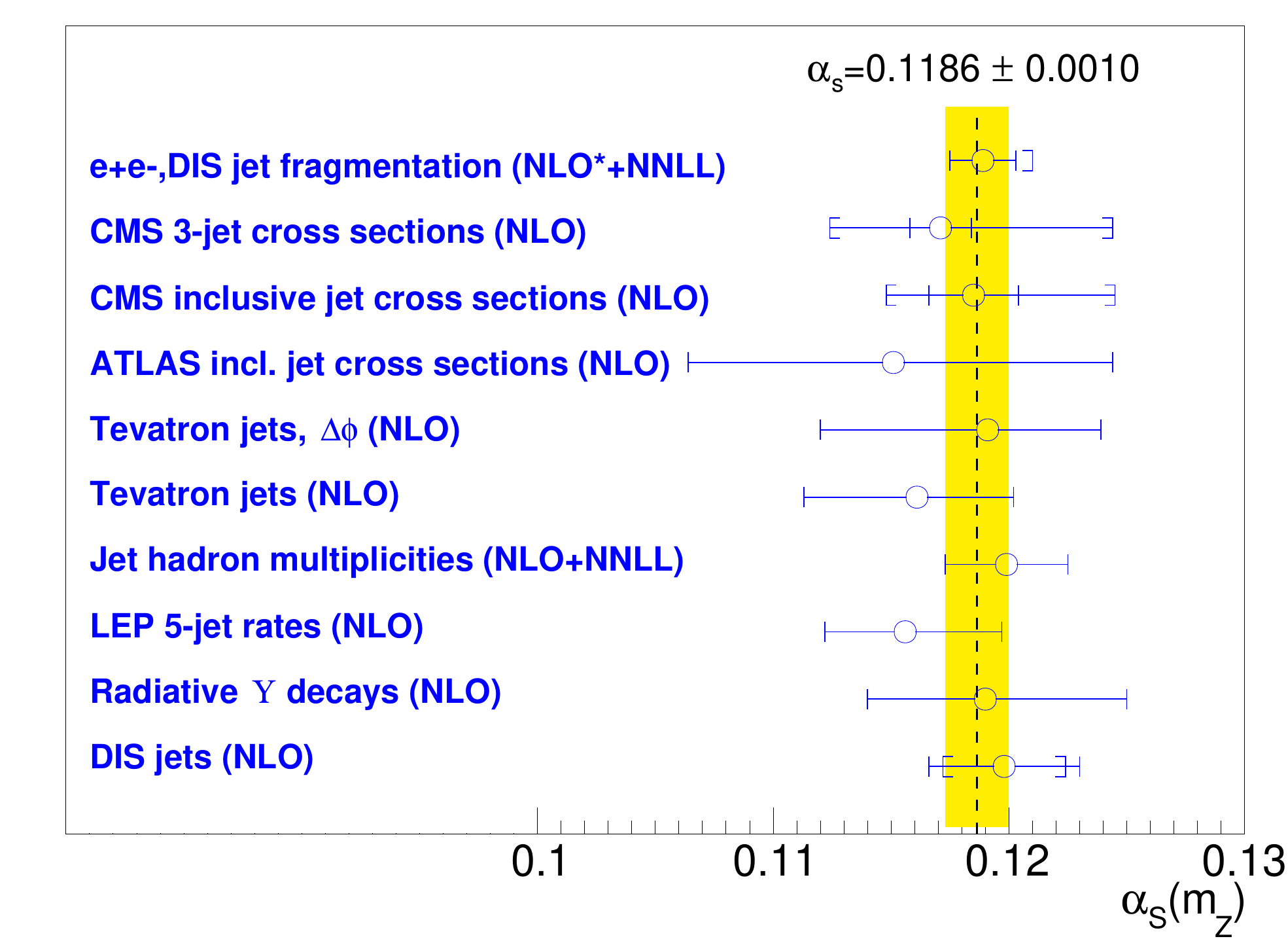}
}
\caption[]{Summary of NLO $\alphas$ determinations using different methods. 
The error ``brackets'' (if present) indicate the theoretical uncertainties of each extraction.
The dashed line and shaded (yellow) band indicate their weighted average (listed also on the top).
}
\label{fig:alphas_NLO}
\end{figure}

In Fig.~\ref{fig:alphas_NLO} we compare our final $\alphasmZ$ value to all other existing results at NLO accuracy
extracted from the latest PDG compilation~\cite{PDG} plus the most recent jet cross sections results
from the CMS~\cite{Khachatryan:2014waa} and ATLAS~\cite{Malaescu:2012ts} data. Our result is the most precise
of all approaches while having a totally different set of experimental and theoretical uncertainties. 
A simple weighted average of all these NLO values yields: $\alphasmZ$~=~0.1186$\;\pm\;$0.0010, in perfect
agreement with the current (NNLO) world-average.

\section{Summary}
\label{sec:conclusion}

The QCD coupling has been determined at NLO*+NNLL accuracy from an analysis of the energy evolution of the
first four moments (multiplicity, peak, width, skewness) of the parton-to-hadron fragmentation functions
measured in $\epem$ and deep-inelastic collisions. A global fit of 360 FF moments in the energy range
$\sqrts\approx$~1--200~GeV, yields $\alphasmZ$~=~0.1189$^{+0.0025}_{-0.0014}$ at the Z mass, in perfect
agreement with the current world-average of $\alphasmZ$~=~0.1185$\;\pm\;$0.0006 (obtained at NNLO accuracy).
The role of higher order corrections determined in this framework for the first time --namely
approximate next-to-leading order (NLO*) fixed-order effects on $\alphas$ plus next-to-next-to-leading log
(NNLL) resummations-- has been assessed by comparing them to older results at LO+NLL accuracy, highlighting
their importance for a quantitative analysis of the FF moments. We have studied in detail the systematic
uncertainties associated with our $\alphas$ extraction, finding that the fit uncertainties obtained through
a $\chi^2$-averaging method fully cover the range of $\alphasmZ$ variations driven by hadronization (finite
hadron-mass) corrections, different experimental final-hadron definitions, as well as the overall fit procedure.  
An additional theoretical-scale uncertainty of +0.0022 has been obtained by redoing the fit of the FF moments
using a different shower cutoff value (i.e. relaxing the limiting-spectrum criterion). The relative
uncertainty of our extracted $\alphasmZ$ value, ($-$1.2\%,$+$2.1\%) is very competitive with respect to the
other methods used so far to determine the QCD coupling.
Work is in progress to include full-NLO (and beyond) corrections
~\cite{DdE_RP} which will further reduce the overall uncertainty (through a possible better data-theory
agreement in the fits and/or a reduced theoretical scale error). The methodology presented here provides a novel 
high-precision approach for the determination of the QCD coupling strength complementary to other existing
jet-based methods --such as jet shapes, and yields and ratios of N-jet production cross sections in $\epem$,
DIS and p-p collisions-- and can be used to reduce the overall final uncertainty of the least well known
interaction coupling in nature.


\paragraph*{\bf Acknowledgments}Redamy~P\'erez-Ramos acknowledges support from the Academy of Finland,
Projects No. 130472 and~133005.

\end{document}